\documentclass[prd,aps,showpacs,secnumarabic,superscriptaddress]{revtex4}
\usepackage[dvips]{graphicx,color}
\begin{document}
\def\ds{\displaystyle}
\def\ss{\scriptstyle}
\def\sb{\mbox{\rule{0pt}{8pt}}}
\def\hh{\mbox{\rule[-6pt]{0pt}{20pt}}}
\title{Modeling dark energy with a top-down approach}
\author{Duane A. Dicus}
\affiliation{Center for Particle Physics, University of Texas, Austin,TX 78712} \email{dicus@physics.utexas.edu}
\author{Wayne W. Repko}
\affiliation{Department of Physics and Astronomy, Michigan State University, East Lansing, MI 48824} \email{repko@pa.msu.edu}
\date{\today}

\begin{abstract}
    We investigate a top-down approach for modeling the dark energy where we fit the luminosity distances directly rather than indirectly fitting the equation of state.
\end{abstract}
\pacs{98.80 -k, 98.80.Es} \maketitle

\section{Introduction}

Since the discovery of dark energy \cite{hiz,super} most of the attempts to understand it have consisted of guessing an equation of state as a function of the redshift parameter, $z$, evaluating one integral over $z$ to determine the dark energy density and then another, necessarily numerical, integral to find a luminosity distance which can be compared with the experimental data \cite{DR}. Thus it is not surprising that there is a lot of degeneracy -- different equations of state give similar or even almost identical fits. What we do here is guess a form for the luminosity distance  and confront our guess with the data straight away.  Then we need only differentiate it once with respect to $z$ to find the dark energy density and once more to find the equation of state. Thus we can, in principle, have analytic expressions for everything.

This is not totally elementary, however, since a best fit to our guess for the luminosity distance, $D_L(z)$, can result in an expression which violates some known property of $D_L(z)$.  These properties are automatically satisfied in the bottom-up approach. For example, the supernova data are bounded from above by the curve for dark energy alone, $\Omega_X=1, D_L(z)=z(1+z)$, and from below by the curve for only matter, $\Omega_M=1, D_L(z)=2\sqrt{1+z}[\sqrt{1+z}-1]$, so it natural to try a fit of the form
\begin{equation}\label{bad}
D_L(z)\,=\,\frac{(1+az)^{n}}{an}[(1+az)^{n}-1] \,.
\end{equation}
The best fit to this, using the data of Ref.\cite{four} and a flat universe, is $a=4.61$ and $n=0.534\pm0.004$.  Unfortunately, this grows faster than $z$ at large $z$ and, as we show below, that is not allowed because the density of dark energy would be negative for some values of $z$.

The direct or top-down approach discussed here was introduced in a paper by Saini, Raychaudhury, Sahni and Starobinsky \cite{SRSS} for a flat universe. A reexamination of the results in this paper is included in what follows. In the next section we give some formalism and derive some conditions that the luminosity distance must satisfy. In Section 3 we make a few guesses for $D_L(z)$, including the model of Ref. \cite{SRSS}, and fit them to the data.  Finally in Sec. 4 we offer some discussion.

\section{Formalism}

The relevant parameters are the matter density and the curvature defined as $\Omega_M\,\equiv\,8\pi\,G\rho_0/3H^2_0$ and \linebreak[4] $\Omega_k\,\equiv\,-k/(R^2_0H^2_0)$ where $\rho_0, H_0$, and $R_0$ are the current values of the matter density, the Hubble parameter, and the scale, and $k=+1,0,$ or $-1$.  With this method $\Omega_M$ is not a parameter which can be fit so, for simplicity, we will fix $\Omega_M\,=\,0.3$. If $\Omega_k$ is greater than or equal to zero then the (dimensionless) luminosity distance is given by
\begin{equation}\label{DL}
D_L(z)\,=\,\frac{1\,+\,z}{\sqrt {\Omega_k}}\,\left\{\sinh\,\int^{z}_{0}\,dz'\,\frac{\sqrt{\Omega_k}\,H_0}{H(z')}\right\}\,,
\end{equation}
where $H(z)$ is the Hubble parameter,
\begin{equation}\label{H}
\frac{H(z)}{H_0}\,=\,\sqrt{\Omega_M\,(1\,+\,z)^3\,+\,\Omega_X\,f(z)\,+\,\Omega_k\,(1\,+\,z)^2\,}\,,
\end{equation}
with $f(z)$ given in terms of the equation of state for dark energy, $p_X\,=\,w(z)\,\rho_X$, as
\begin{equation}\label{f}
f(z)\,=\,\exp\,\left\{\,3\,\int^{z}_{0}\,\frac{dx}{1\,+\,x}\,(1\,+\,w(x))\,\right\}\,,
\end{equation}
and
\begin{equation}\label{omega}
\Omega_X\,=\,1\,-\,\Omega_M\,-\,\Omega_k\,.
\end{equation}
If $\Omega_{k}$ is less than zero the the hyperbolic sine in Eq.(\ref{DL}) is replaced by the ordinary sine and $\Omega_{k}$ is replaced by $|\Omega_{k}|=-\Omega_{k}$.  However, after we take one derivative our expressions will be valid for all $\Omega_{k}$. Note that $f(z)\,>\,0$ for all $z$ and $f(0)\,=\,1$.

The simplest and most obvious of the conditions on $D_L(z)$ is that it must be positive for all $z$.  Also, for $z<<1$, $D_L(z)$ must equal $z$,
\begin{eqnarray}
D_L(z)\,&>&\,0,   \label{pos}  \\
D_L(z)\,&=&\,z,  \hspace{1cm} z<<1.  \label{lowz}
\end{eqnarray}

Now to go further we need to solve Eq.(\ref{DL}) for $f(z)$,  so rewrite it as
\begin{equation}\label{der1}
\left[\,\sinh^{-1}\left(\,\frac{\sqrt{\,\Omega_k}\,D_L(z)}{1\,+\,z}\,\right)\right]^{'} \,=\,\frac{\sqrt{\,\Omega_k}\,H_0}{H(z)}\,,
\end{equation}
where the prime indicates a derivative with respect to $z$.  Differentiating the $\sinh^{-1}$ gives
\begin{equation}\label{der2}
\left[\frac{D_L(z)}{1\,+\,z}\right]^{'}\,\frac{1\,+\,z}{\sqrt{(1+z)^2+\Omega_k\,D_L^2(z)}}\,= \,\frac{H_0}{H(z)}\,,
\end{equation}
or, finally, using Eq.\,(\ref{H}),
\begin{equation}\label{der3}
\frac{(1\,+\,z)\,D_L^{'}(z)\,-\,D_L(z)}{\sqrt{(1+z)^2\,+\,\Omega_k\,D_L^2(z)}}\, =\,\frac{1\,+\,z}{\sqrt{\Omega_M(1+z)^3\,+\,(1-\Omega_M-\Omega_k)\,f(z)\,+\,\Omega_k\,(1+z)^2}} \,.
\end{equation}
Eq.(\ref{der3}) is valid for all $\Omega_{k}$.

Introducing $D_1(z)$ as
\begin{equation}\label{D1}
D_{1}(z)\,\equiv\,(1+z)\,D^{'}_L(z)\,-\,D_L(z)\,,
\end{equation}
we note that, from Eq.\,(\ref{der3}), $D_1(z)$ must be positive. This is not always easy to satisfy. Solving Eq.(\ref{der3}) for $f(z)$ gives
\begin{equation}\label{feq}
f(z)\,=\,\frac{(1+z)^2}{1\,-\,\Omega_M\,-\Omega_k}\,\left\{\frac{(1+z)^2\, +\,\Omega_k\,D_L^2(z)}{D^2_1(z)}\,-\,\Omega_M\,(1+z)\,-\,\Omega_k\,\right\}\,.
\end{equation}
Since $f(z)$ must be positive the terms in the $\{\,\,\}$ brackets must be positive. On the other hand, for $z\,>>\,1$, we want the energy density in dark energy to be less than that of matter, i.e.,
\begin{equation}\label{zbt1}
(1\,-\,\Omega_M\,-\,\Omega_k)\,f(z)\,<\,\Omega_M\,(1+z)^3.
\end{equation}
So Eq.(\ref{feq}) gives
\begin{equation}\label{allz}
\Omega_M\,(1+z)\,+\,\Omega_k\,<\,\frac{(1+z)^2\,+\,\Omega_k\,D^2_L(z)}{D_1^2(z)}\,, \qquad\mbox{all}\,\,\,z\,,
\end{equation}
and, from Eq.(\ref{zbt1}) we have
\begin{equation}\label{zbig}
\frac{(1+z)^2\,+\,\Omega_k\,D^2_L(z)}{D_1^2(z)}\,<\,2\,\Omega_M(1+z)\,+\,\Omega_k\,, \hspace{1cm}z\,>>\,1\,.
\end{equation}

These conditions can be somewhat simplified by writing Eq.(\ref{DL}) for $D_L(z)$ as
\begin{equation}\label{DLY}
D_L(z)\,=\,\frac{1+z}{\sqrt{\,\Omega_k}}\,\sinh\left(\frac{\sqrt{\,\Omega_k}\,Y(z)}{1+z}\right)\,,
\end{equation}
where $D_L(z)\,=\,Y(z)$ if $\Omega_k$ is zero.  Then $f(z)$ of Eq.\,(\ref{feq}) reduces to
\begin{equation}\label{fY}
f(z)\,=\,\frac{(1+z)^2}{1-\Omega_M-\Omega_k}\,\left\{\,\frac{(1+z)^2}{D_Y^2(z)} \,-\,\Omega_M\,(1+z)\,-\,\Omega_k\,\right\}\,,
\end{equation}
with
\begin{equation}\label{DY}
D_Y(z)\,\equiv\,(1+z)\,Y^{'}(z)\,-\,Y(z)\,.
\end{equation}
$D_1(z)$, given by Eq.(\ref{D1}), is proportional to $D_Y(z)$ and thus $D_Y(z)$ must be positive.

The constraint equations (\ref{allz}) and (\ref{zbig}) become
\begin{equation}\label{allbigz}
\Omega_M\,(1+z)\,+\,\Omega_k\,<\,\frac{(1+z)^2}{D_Y^2(z)}\,<\,2\,\Omega_M\,(1+z)\,+\,\Omega_k\,,
\end{equation}
where the first inequality holds for all $z$ and the second is valid only for $z>>1$. For $z$ large the only solution which satisfies this and has $Y(z)$ positive is
\begin{equation}\label{Ybigz}
Y(z)\,\sim\,Az\,-\,B\sqrt{z}
\end{equation}
with $A$ and $B$ positive constants. The $A$ term, whose leading $z$ behavior cancels in $D_{Y}(z)$, is necessary to keep $D_{Y}(z)$ positive. For the cosmological constant case, where $f(z)=1$ for all $z$, $B$ equals $2/\sqrt{\Omega_{M}}$. For small $z$, $Y(z)$ must, of course, go as $z$. Then, to leading order in $z$, Eq.(\ref{allbigz}) becomes
\begin{equation}\label{approx}
\Omega_M\,<\,\xi\,<\,2\,\Omega_M
\end{equation}
with
\begin{equation}\label{X}
\xi\,=\,\frac{4}{B^2}\,.
\end{equation}
Condition (\ref{approx}) is not perfect because it is only valid to leading order in $\sqrt{z}$ which is at most $\sqrt{z}=\sqrt{1089}= 33$. In some models, such as Eq.\,(\ref{todd}) in the next section, the next term in Eq.\,(\ref{Ybigz}), the constant term, is large and making it impossible to satisfy Eq.\,(\ref{allbigz}). Consequently, condition (\ref{approx}) is a necessary but not a sufficient condition and for each fit to the data it is still necessary to check that the conditions in Eq.\,(\ref{allbigz}) are satisfied. Eq.\,(\ref{approx}) is only a guide as to where to look for a solution.

Once we find a model which satisfies these constraints, the dark energy equation of state is given by
\begin{equation}\label{w}
w(z)\,=\,-1\,+\,\frac{1}{3}(1+z)\frac{f'(z)}{f(z)}\,,
\end{equation}
which, using Eq.\,(\ref{fY}), can be written as
\begin{equation}\label{wp}
w(z)\,=\,\frac{1}{3\,D_Y(z)}\frac{(1+z)^2D_Y(z)+\Omega_kD_Y^3(z)-2(1+z)^4Y''(z)} {(1+z)^2-(1+z)\Omega_MD_Y^2(z)-\Omega_kD_Y^2(z)}\,.
\end{equation}
Note that if we wish to satisfy the condition that $w(z)>-1$ then $f'(z)>0$ i.e., $\rho_{X}$ must increase with $z$ (just not as fast as $\rho_M$).  We will not worry about restricting $w(z)$ in this paper.

\section{Models}

For the data we use the ``gold'' set of Ref.\cite{four} plus the seven events given in Ref.\cite{essence} the total of which we will call the ``essence'' data.  We also use the data set of Ref.\cite{legacy} which we call the ``legacy'' data.  We compare these data sets with $5\log_{10}D_L(z)\,+\,M$ where $M$ is an offset we fit as an extra parameter. WMAP \cite{WMAP} gives a comoving distance to the decoupling surface and we translate this into a (dimensionless) luminosity distance and add it to whichever of the two data sets we are using \cite{WT}. Specifically, for this point we take $5\log_{10}D_L(z)=17.79\pm0.11$ at $z=1089$ and we add it with no offset.

From the previous section what is needed are expressions for $Y(z)$ which give $D_L(z)\,=\,z$ for $z<<1$, behave as Eq.\,(\ref{Ybigz}) for $z>>1$, and have $D_Y(z)\,>\,0$ for all $z$.  The model which most closely resembles the form discussed in the Introduction but also satisfies these conditions is
\begin{equation}\label{Yss}
Y(z)\,=\,\frac{2}{c}\,\sqrt{1+az}\,[\sqrt{1+cz}\,-\,1]
\end{equation}
with
\begin{equation}\label{ca}
\xi\,=\,\frac{c^2}{a}\,.
\end{equation}
For $\Omega_M\,=\,0.3$ the best fit to the essence data gives $\xi\,=\,0.6$, $a\,=\,2.74$, $\Omega_k\,=\,0.112$ and $\chi^2$ of 187.1 but, for large $z$, the ratio of energy densities $\rho_X/\rho_M$ is much too large, $\sim0.9$. If we fix $\xi$ at $0.35$ a fit gives $\chi^2\,=\,187.7$, $a\,=\,2.43\pm0.20$, $\Omega_k\,=\,0.029\pm0.036$. In
\begin{figure}[h]
\begin{minipage}[t]{0.45\textwidth}
  \centering
  \includegraphics[width=3.0in]{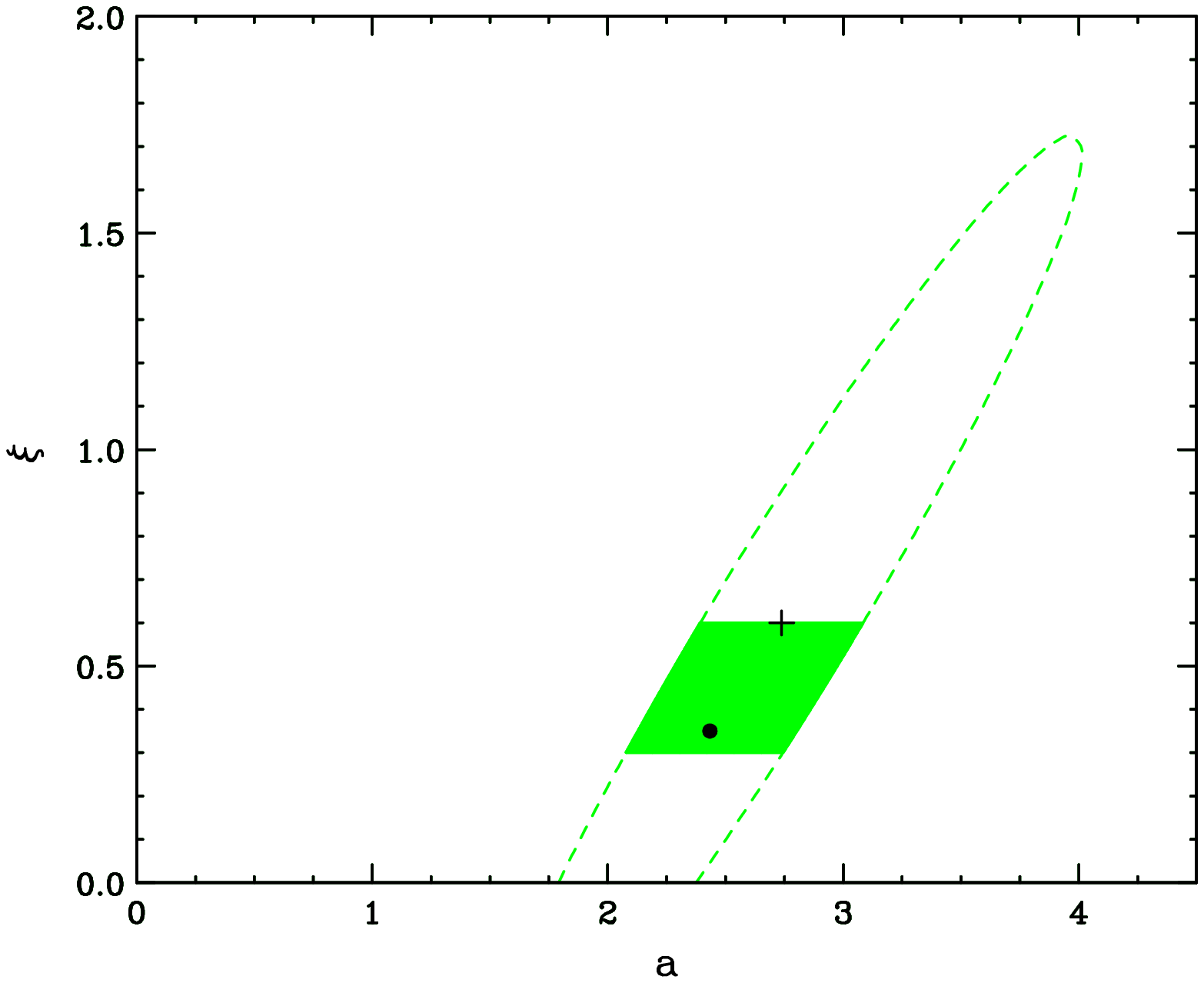}
  \caption{The dashed curve is $68.3\%$ confidence contour for the $Y(z)$ of Eq.\,(\ref{Yss}) in the $a\,-\,\xi$ using the essence data. Points in the shaded area satisfy the constraint $\Omega_M\leq \xi\leq 2\Omega_M$. The cross is the minimum of $\chi^2$ and the dot is the point $a=2.43$, $\xi=0.35$.\label{td2contour}}
\end{minipage}%
\begin{minipage}[t]{0.1\textwidth}
\hfil
\end{minipage}%
\begin{minipage}[t]{0.45\textwidth}
  \centering
  \includegraphics[width=3.0in]{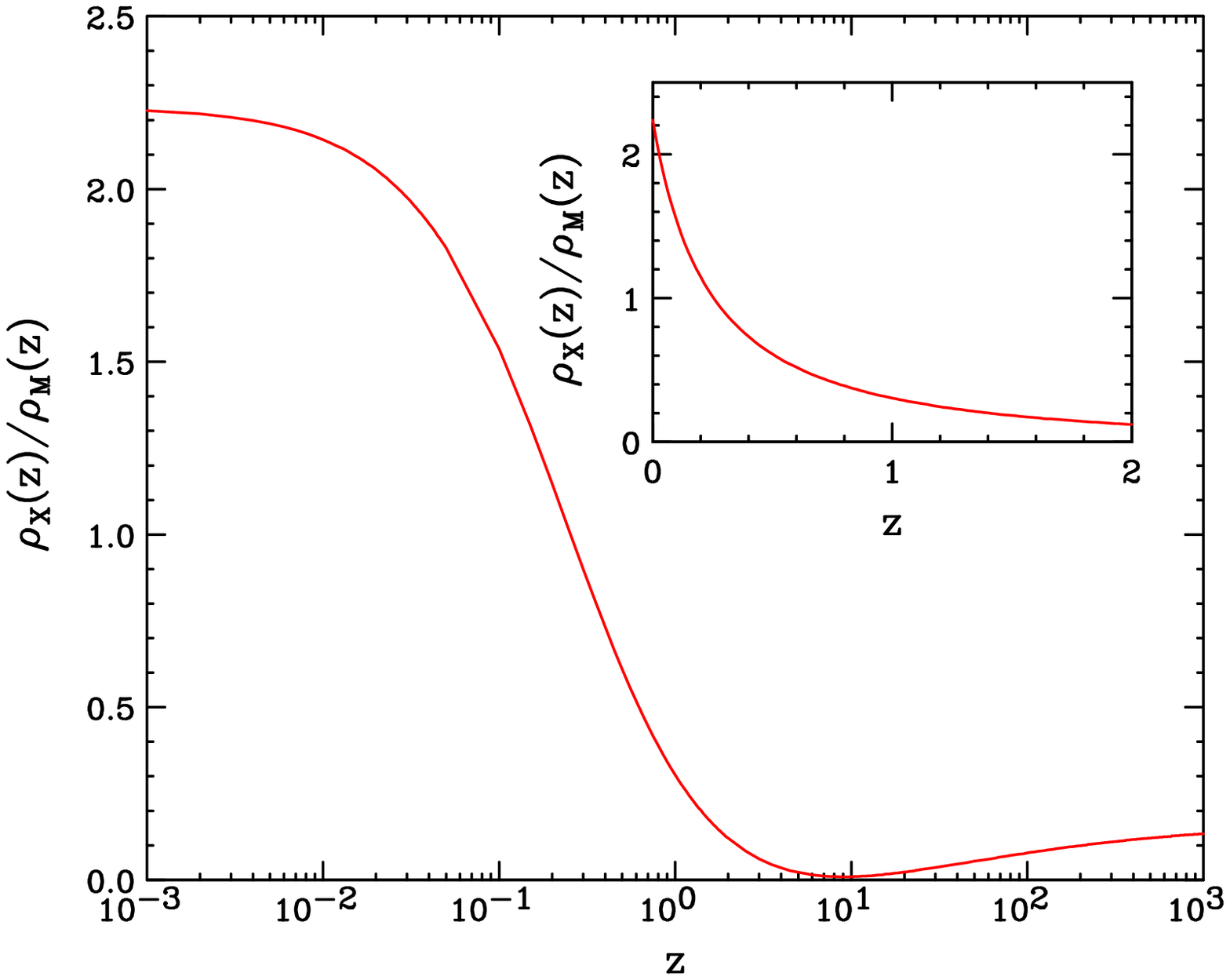}
  \caption{The behavior of the ratio $\rho_X(z)/\rho_M(z)$ is shown for the $Y(z)$ of Eq.\,(\ref{Yss}) using the essence data with $\xi=0.35$ and $\Omega_M=0.3$.\label{td2ratio}}
\end{minipage}
\end{figure}
Fig.\,\ref{td2contour} we show the $68.3\%$ confidence contour in the $a\,-\,\xi$ plane together with the best fit and the $a=2.43$, $\xi=0.35$ point. The latter gives a somewhat better ratio of energy densities at large $z$ of $0.13$, as shown in Fig.\,\ref{td2ratio}. For $\xi$ below $0.35$, $\rho_X$ ($f(z)$) is not positive for all $z$. The legacy data with $\xi\,=\,0.36$ gives $\chi^2=113.0$, $a\,=\,2.50\pm0.18$ and $\Omega_{k}\,=\,0.033\pm0.035$. The ratio of energy densities at large $z$ is $0.16$. Thus the two data sets give essentially the same answer for the parameters and results consistent with $\Omega_k\,=\,0$.

Another expression for $Y(z)$ which satifies the condition (\ref{Ybigz}) is
\begin{equation}\label{Yrn}
Y(z)\,=\,z\,\left[\frac{1\,+\,a\sqrt{z}}{1\,+\,c\sqrt{z}}\right]^{n}\,.
\end{equation}
Here $\xi$ of Eq.(\ref{approx}) is given by
\begin{equation}\label{XX}
\xi\,=\,\frac{4}{n^2}\left(\frac{c}{a}\right)^{2n}\frac{a^2c^2}{(a-c)^2}\,.
\end{equation}
\begin{figure}[h]
\begin{minipage}[t]{0.45\textwidth}
  \centering
  \includegraphics[width=3.0in]{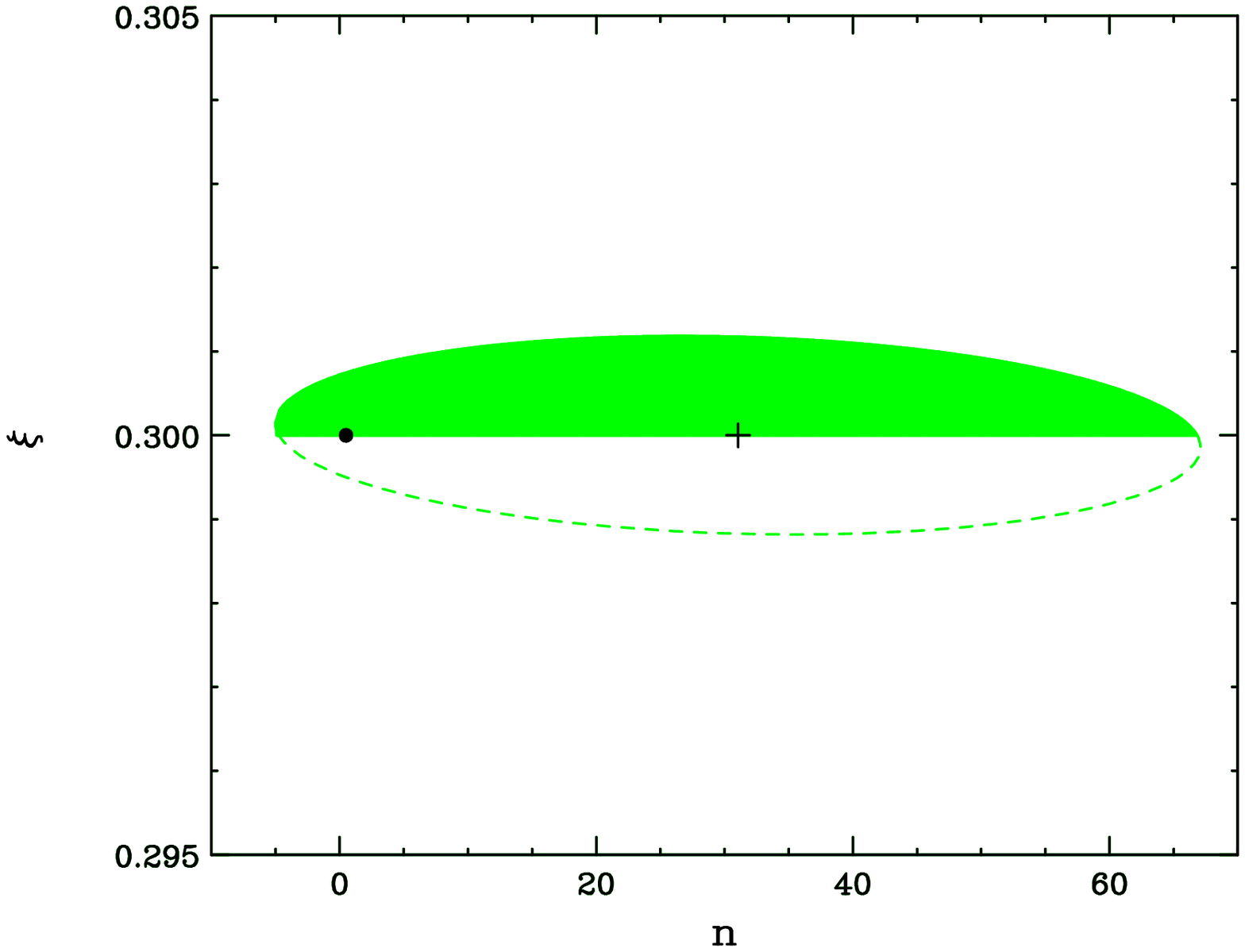}
  \caption{The $68.3\%$ confidence contour in the $n\,-\,\xi$ plane is shown for the $Y(z)$ of Eq.\,(\ref{Yrn}) using the essence data. Points in the shaded area satisfy the constraint $\Omega_M\leq \xi\leq 2\Omega_M$.  The cross is the minimum of $\chi^2$ and the dot is the point $n=0.5$, $\xi=0.30$. \label{td1contour}}
\end{minipage}%
\begin{minipage}[t]{0.1\textwidth}
\hfil
\end{minipage}%
\begin{minipage}[t]{0.45\textwidth}
  \centering
  \includegraphics[width=3.0in]{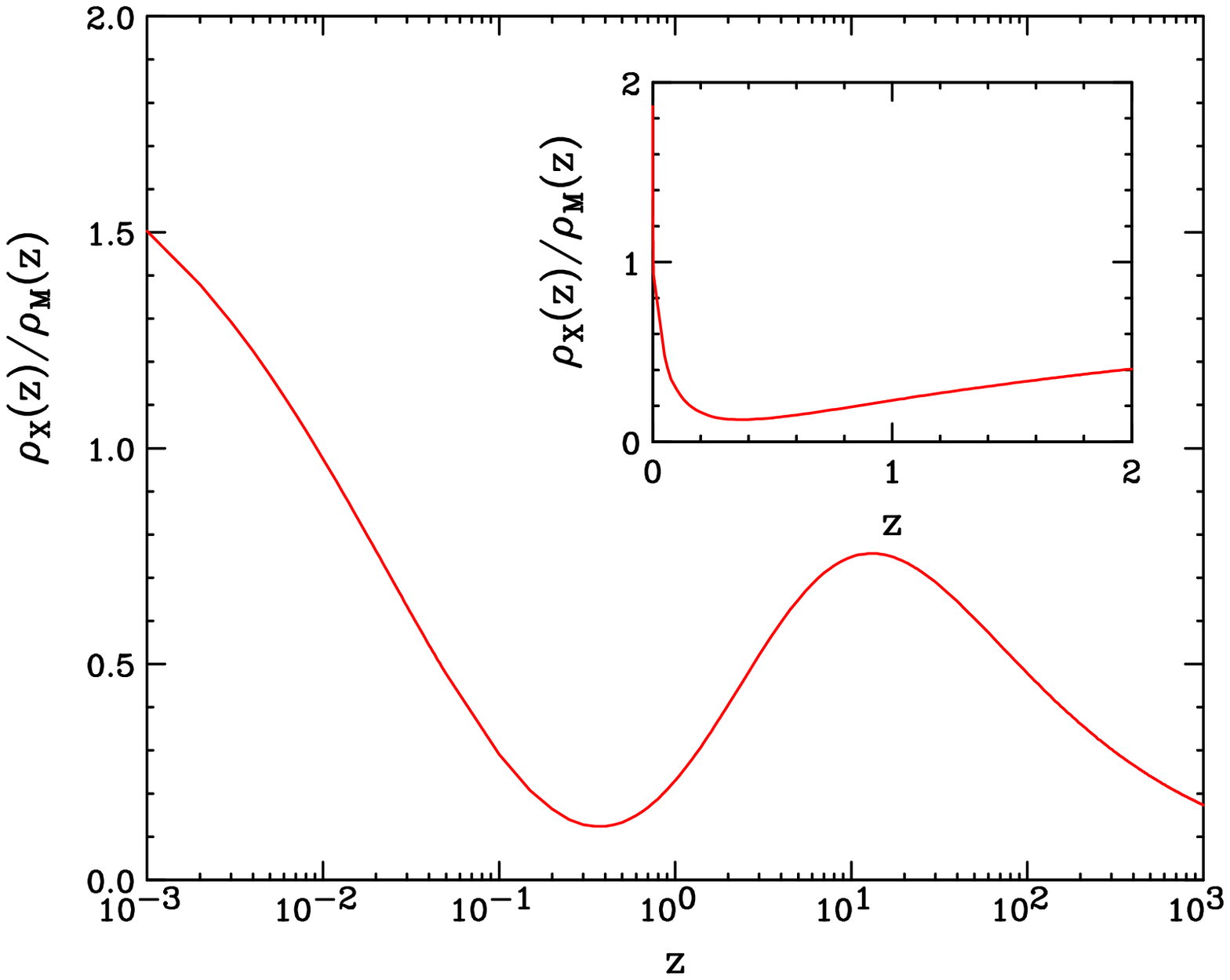}
  \caption{The behavior of the ratio $\rho_X(z)/\rho_M(z)$ is shown for the $Y(z)$ of Eq.\,(\ref{Yrn}) using the essence data with $\xi=0.30$ and $\Omega_M=0.3$.\label{td1ratio}}
\end{minipage}
\end{figure}
Unlike the model given in Eq.\,(\ref{Yss}), the ratio of energy densities $\rho_X/\rho_M$ seems to be positive for all values of $\xi$ which satisfy condition\,(\ref{approx}) but, like the case of Eq.\,(\ref{Yss}), this ratio is never very small. It turns out that the minimum $\chi^2$ occurs when $a\,\approx\,c$. Then to satisfy condition (\ref{approx}) together with Eq.\,(\ref{XX}) requires $c\,<\,a$, which, in turn, forces $n$ to be very large in order to cancel the small denominator in Eq.\,(\ref{XX}).  This gives a lot of degeneracy. For example, using the essence data, the minimum in $\chi^2$ occurs at $185.7$ when the $\xi$ parameter is 0.3, the other parameters are $a=0.918$, $c=0.900$, $n=34.57$, $\Omega_k=0.104$ and the ratio of energy densities at large $z$ is $0.11$. But if we allow $\chi^2$ to increase by one we can leave $\xi$ unchanged, have $n=\frac{1}{2}$ and $a=2.97$, $c=0.351$, $\Omega_k=0.141$ as shown in Fig.\,\ref{td1contour}. In this case, the ratio of energy densities  at high $z$ is $0.17$ and its behavior is illustrated in Fig.\,\ref{td1ratio}. The legacy data gives very similar results although allowing $\chi^2$ to change by one only gets $n$ down to $0.9$ where $a=1.68$ and $c=0.504$. The only surprise is that the $\chi^2$ is larger than expected at $123.8$.

A brute force expression for $Y(z)$ is
\begin{equation}\label{todd}
Y(z)\,=\,A\,\frac{z^2}{a+z}\,-\,2\,B\,\frac{z^2}{b+z^{3/2}}\,+\,\frac{z}{1+cz^{n}}\,,
\end{equation}
with $A$, $a$, $B$, $b$, and $c$ greater than zero and $n$ greater than $\frac{1}{2}$. At large $z$ the first and second terms on the right-hand-side reproduce Eq.\,(\ref{Ybigz}) while at small $z$ the last term gives the desired linear dependence on $z$.  But this expression doesn't seem to work at all. A fit to the data gives seemingly reasonable values for $\chi^2$ and for the parameters;  for example, using the essence data we find $A=4.80, a=2.95, B=1.8, b=35.0, n=1.92, c=1.28$ and $\Omega_k=-0.0883$ with $\chi^2=184.5$. But the omitted constant term of Eq.(\ref{Ybigz}) is equal to $A(1+a)$ which is large enough that the lower bound in Eq.\,(\ref{allbigz}) is always violated for some values of $z$.

Finally we consider the model given in Ref.\cite{SRSS} which in our notation is
\begin{equation}\label{S4}
Y(z)\,=\,2(1+z)\left[\frac{1+z-\alpha\sqrt{1+z}-1+\alpha} {\beta(1+z)+\gamma\sqrt{1+z}+2-\alpha-\beta-\gamma}\right]\,,
\end{equation}
with
\begin{equation}\label{xi}
\xi\,=\,\left(\frac{\beta^2}{\alpha\beta+\gamma}\right)^2\,.
\end{equation}
For the essence data the minimum $\chi^2$ occurs at
\begin{eqnarray}
\alpha &=&1.93^{+0.03}_{-0.12} \hspace{76pt} \beta\,=\,0.66^{+0.09}_{-0.03}  \nonumber  \\
\gamma &=&\-0.51^{+0.06}_{-0.19} \hspace{70pt} \Omega_k\,=\,0.09^{+0.19}_{-0.05}  \label{num}
\end{eqnarray}
with $\chi^2\,=\,184.8$.  This $\chi^2$ is smaller than those obtained for the models above, in particular model (\ref{Yss}), but this probably isn't particularly significant; for $\Omega_k$ equal to zero $\chi^2$ jumps to 188.4. The ratio of energy densities at large $z$ is 0.12 which is about the same as model (\ref{Yss}). If we push the parameters to the lower bounds given by the errors the energy density ratio gets smaller but at the upper bounds it is unreasonably high ($\sim 1$).  The legacy data gives a reasonable fit with $\chi^2=112.5$, but the central values of the parameters give a dark energy density which is negative for $2\le\,z\le\,40$. Here, too, there are values of the parameters within the errors that result in a positive dark energy density for all $z$.

\section{Discussion}

In Sec. 2 we derived some necessary conditions to be used as a guide in implementing this top-down approach for studying the dark energy.  The simple models in Sec. 3 show that these conditions are not always sufficient. The model given by Eq.\,(\ref{todd}) satisfies the conditions but has no solution for the parameters which works for all $z$. The model given by Eq.\,(\ref{Yrn}) is  worse in the sense that it has satisfactory solutions for an extremely wide range of values of the parameters. Knowing that a model does not work is useful since we can proceed to other forms. Having a model which works but where we cannot gainfully limit the parameters leaves us in a sort of limbo. In short the problem of degeneracy so prevalent in the bottom-up approach is no less severe here.  In addition this model has a large bump in the ratio of the energy densities around $z=10$ (see Fig.\,\ref{td1ratio}) which seems unlikely to be physical.

Thus we are left with the models of Eq.(\ref{Yss}) or Eq.(\ref{S4}). For model (\ref{Yss}) the best fit has an energy density ratio, $\rho_X/\rho_M$, which does not go to zero at large $z$ but we can adjust the parameters to find a model where the ratio does get small without too much penalty in $\chi^2$.  This is the case shown in Figs.\,\ref{td2contour} and \ref{td2ratio}. Even then the model is not without problems. The equation of state $w(z)$, as calculated from Eq.\,(\ref{wp}), is less than $-1$ at $z=0$, but we have agreed not to worry about conditions on $w(z)$ here; even in a bottom-up approach it is not always possible to find fits to the data satisfying $w(z)>-1$ for all $z$.

In some ways the model given by Eq.\,(\ref{S4}) is the most satisfactory. It can have a small $\chi^2$ if we allow $\Omega_k$ to be nonzero and it can have small values for the ratio of dark energy to matter densities at large redshift if we allow the parameters to be one sigma off the minimum in the most favorable direction. However, if the parameters shift in the other direction the model rapidly becomes very unsatisfactory. Also here again $w(0)$ is less than $-1$, sometimes much less.

On the positive side the data seem well behaved. For other reasons we think that $\Omega_k$ must be zero so it is pleasing to see that both data sets are consistent with this for the cases that we tried. The predictions for the parameters were also very similar for the two sets.  However, a big part of this agreement may come from our inclusion of the WMAP point which has a strong normalizing effect.

In summary it seems very hard to guess a form for the luminosity distance, $D_L(z)$, which gives results that are physically sensible. This is not a problem in the bottom-up method. There it is all too easy to guess forms for $w(z)$ which are sensible physically and automatically satisfy the conditions given above. Also the analytic solutions for the dark energy equation of state, which certainly exist in this top-down method, and are one of the main reasons for this approach, are likely to be rather complicated (see Eq.\,(\ref{wp})). So it is hard to know whether, with more data, this \nopagebreak[0] scheme of directly fitting the luminosity distances will be useful.

\begin{acknowledgments}
We thank Yun Wang for a clarifying discussion about the comoving distance measured by WMAP and Varun Sahni for drawing our attention to Ref.\cite{SRSS}. This work was supported in part by the U.~S. Department of Energy under Grant No. DE-FG03-93ER40757 and by the National Science Foundation under Grant PHY-0244789.
\end{acknowledgments}

\end{document}